\documentclass[pra,floatfix,twocolumn,superscriptaddress,showpacs,preprintnumbers,longbibliography,nofootinbib]{revtex4-1}

\usepackage[utf8]{inputenc} 
\usepackage{bibunits} 
\usepackage{amssymb,amsfonts,amsmath}
\usepackage{float} 
\usepackage{amsthm}
\usepackage{enumitem}
\usepackage{bbold}
\usepackage[normalem]{ulem}
\usepackage{hyperref} 
\hypersetup{colorlinks=true, urlcolor=blue, citecolor=blue,linkcolor=blue}
\usepackage[all]{hypcap} 

\usepackage{graphicx}
\usepackage{dcolumn}
\usepackage{bm}
\usepackage{psfrag}
\usepackage{epsfig}
\usepackage{multirow}
\usepackage{color}
\usepackage{bbm}
\usepackage[FIGTOPCAP,raggedright,nooneline]{subfigure}
\usepackage{verbatim}
\usepackage{upgreek} 
\usepackage{physics}
\usepackage{tikz} 

\newcommand{\T}[1]{\text{#1}}

\newcommand{\ignore}[1]{}

\newcommand{\eq}{Eq.\,}
\newcommand{\eqs}{Eqs.\,}
\newcommand{\fig}{Fig.\,}
\newcommand{\figs}{Figs.\,}
\newcommand{\cf} {cf.~}

\newcommand{\ie} {i.e.~}
\newcommand{\eg} {e.g.~}
\newcommand{\rref} {Ref.\,}
\newcommand{\rrefs} {Refs.\,}

\newcommand{\Lc}{\kappa}

\newcommand{\Det}{ \textrm{Det}}

\newcommand{\Hk}{\tilde{H}}
\newcommand{\G}{\hat{G}}
\newcommand{\res}[1]{\underset{#1}{\rm{Res}}}

\begin{document}

\title{Exotic interactions mediated by a non-Hermitian photonic bath}

\author{Federico Roccati}
\affiliation{Universit$\grave{a}$  degli Studi di Palermo, Dipartimento di Fisica e Chimica -- Emilio Segr$\grave{e}$, via Archirafi 36, I-90123 Palermo, Italy}

\author{Salvatore Lorenzo}
\affiliation{Universit$\grave{a}$  degli Studi di Palermo, Dipartimento di Fisica e Chimica -- Emilio Segr$\grave{e}$, via Archirafi 36, I-90123 Palermo, Italy}

\author{Giuseppe Calajò}
\affiliation{ICFO-Institut de Ciencies Fotoniques, The Barcelona Institute of Science and Technology, 08860 Castelldefels (Barcelona), Spain}

\author{\\G. Massimo Palma}
\affiliation{Universit$\grave{a}$  degli Studi di Palermo, Dipartimento di Fisica e Chimica -- Emilio Segr$\grave{e}$, via Archirafi 36, I-90123 Palermo, Italy}
\affiliation{NEST, Istituto Nanoscienze-CNR, Piazza S. Silvestro 12, 56127 Pisa, Italy}

\author{Angelo Carollo}
\affiliation{Universit$\grave{a}$  degli Studi di Palermo, Dipartimento di Fisica e Chimica -- Emilio Segr$\grave{e}$, via Archirafi 36, I-90123 Palermo, Italy}
\affiliation{Radiophysics Department, National Research Lobachevsky State University of Nizhni Novgorod, 23 Gagarin Avenue, Nizhni Novgorod 603950, Russia}

\author{Francesco Ciccarello}
\affiliation{Universit$\grave{a}$  degli Studi di Palermo, Dipartimento di Fisica e Chimica -- Emilio Segr$\grave{e}$, via Archirafi 36, I-90123 Palermo, Italy}
\affiliation{NEST, Istituto Nanoscienze-CNR, Piazza S. Silvestro 12, 56127 Pisa, Italy}

		\date{\today}
	
		\begin{abstract}
			Photon-mediated interactions between quantum emitters in engineered photonic baths is an emerging area of quantum optics. At the same time, non-Hermitian (NH) physics is currently thriving, spurred by the exciting possibility to access new physics in systems ruled by non-trivial NH Hamiltonians - in particular photonic lattices - which can challenge longstanding tenets such as the Bloch theory of bands. Here, we combine these two fields and study the exotic interaction between emitters mediated by the photonic modes of a lossy photonic lattice described by a NH Hamiltonian. We show in a paradigmatic case study that {\it structured losses} in the field can seed exotic emission properties. Photons can mediate dissipative, fully non-reciprocal, interactions between the emitters with range critically dependent on the loss rate. When this loss rate corresponds to a bare-lattice exceptional point, the effective couplings are exactly nearest-neighbour, implementing a dissipative, fully non-reciprocal, Hatano-Nelson model. Counter-intuitively, this occurs irrespective of the lattice boundary conditions. Thus photons can mediate an effective emitters’ Hamiltonian which is translationally-invariant despite the fact that the field is not. We interpret these effects in terms of metastable atom-photon dressed states, which can be exactly localized on only two lattice cells or extended across the entire lattice. These findings introduce a new paradigm of light-mediated interactions with unprecedented features such as non-reciprocity, non-trivial dependence on the field boundary conditions and range tunability via a loss rate.
		\end{abstract}

		\maketitle

		\section{Introduction}
		The irreversible leakage of energy into an external reservoir is traditionally viewed as a detriment in physics as losses usually spoil the visibility of several phenomena, in particular those relying on quantum coherence. A longstanding tool for describing these detrimental effects are non-Hermitian (NH) Hamiltonians \cite{CohenAP}. While their introduction dates back to the early age of quantum mechanics \cite{gamow1928quantentheorie}, only in recent years it was realized and experimentally confirmed that systems described by NH Hamiltonians can exhibit under suitable conditions a variety of exotic phenomena  \cite{el-ganainyNP2018,UedaReview}. Among these are: coalescence of eigenstates at exceptional points \cite{miri2019exceptional}, unconventional geometric phase \cite{GPPRE}, failure of bulk-edge correspondence \cite{LeePRL2016}, critical behavior of quantum correlations around exceptional points \cite{caoPRL2020,huber2020emergence,roccati2021quantum}, non-Hermitian skin effect \cite{YaoPRL2018}. As a typical consequence, traditional tenets of Physics such as the Bloch theory of bands and even the very notion of ``bulk" may require a non-trivial revision in the non-Hermitian realm \cite{BlochBH}.
		Such NH effects are intensively studied in several scenarios (such as mechanics, acoustics, electrical circuits, biological systems) \cite{UedaReview} and, most notably in view of our purposes here, optics and photonics \cite{fengNP2017,longhiEPL2017}. 
		
		Here, we investigate NH physics in a setup comprising a set of emitters (such as atoms, superconducting qubits or resonators) coupled to a photonic lattice, implemented \eg by an array of coupled cavities \cite{ShamoomPRA2013,Lombardo2014b,Douglas2015b,Calajo2016b,Tao2016,Gonzalez-Tudela2017a,Gonzalez-Tudela2017b,Gonzalez-Tudela2018a,Tudela2018Quantum,Tudela2018anisotropic,Sanchez-Burillo2019a,LeonfortePRL2021,PeterHall21}. 	
		Such type of systems are currently spurring considerable interest in the quantum optics community in particular due to the possibility of tailoring directional emission \cite{RamosPRA16,Gonzalez-Tudela2017a,ZuecoSaw2020} or exploiting photon-mediated interactions between the emitters to engineer effective spin Hamiltonians \cite{Douglas2015b,Gonzalez-Tudela2017b,Gonzalez-Tudela2018a,LeonfortePRL2021, KockumPRLTopo21}. Remarkably, the range and profile of these second-order interactions are directly inherited from the form of atom-photon dressed states (typically arising within photonic bandgaps) which in turn depends on the lattice {\it structure} \cite{Lambropoulos2000a}.  
		Experimental implementations were demonstrated in various architectures such as circuit QED \cite{Liu2017a,Sundaresan2019,scigliuzzo2021probing}, cold atoms coupled to photonic crystal waveguides \cite{KimblePNAS2016} and optical lattices \cite{Krinner2018,Stewart2020}.

		Studying the spoiling effect of photon leakage in such quantum optics setups is a routine task, even through NH Hamiltonians (see \eg \rref\cite{Calajo2016b}), the usual configuration considered being yet that of uniform losses. In contrast, here we introduce an engineered {\it pattern} of photonic losses so as to affect the photonic normal modes. The basic question we ask is whether and to what extent shaping the field structure through patterned leakage  (besides photonic hopping rates) can affect the nature of atom-photon interactions, hence photon-mediated couplings. 
		
		By considering a paradigmatic case study, we will in particular show that photons can mediate dissipative non-reciprocal interactions between the emitters with exotic features such as: (i)loss-dependent interaction range (from purely long-range to purely nearest-neighbour), (ii) formation of short- and long-range metastable dressed states and (iii) insensitivity to the field boundary conditions (BCs).

		\section{Setup and Hamiltonian}
\begin{figure}[t]
			\centering
			\includegraphics[width=8.5cm]{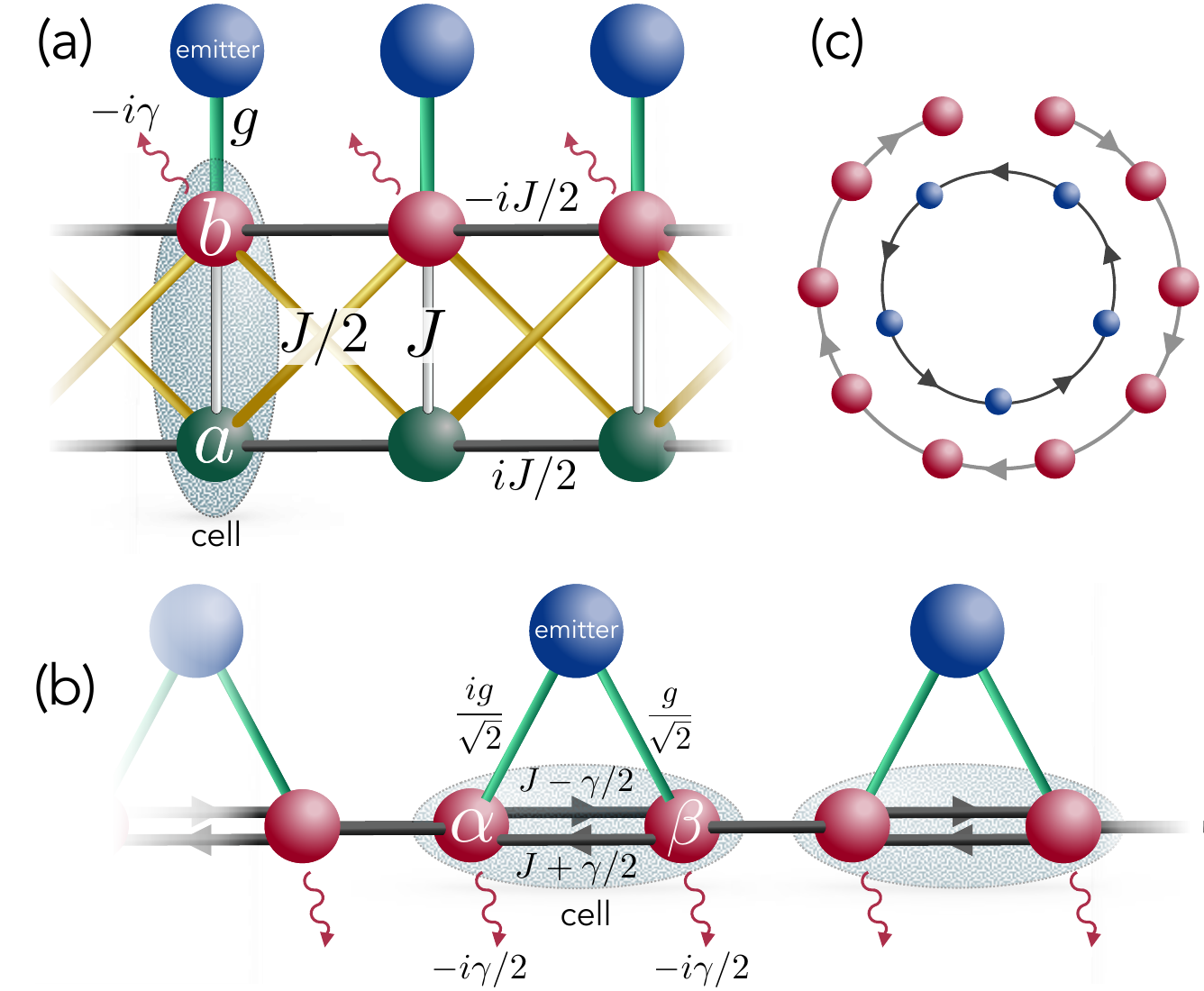}
			\caption{(a): Setup: photonic lattice with unit cell comprising a pair of cavities labeled $a$ (lossless) and $b$ (lossy). Each quantum emitter is locally coupled to a lossy cavity. 
				(b): Same setup as (a) in the picture defined by the unitary (intra-cell) transformation \eqref{transformation}. All cavities are now lossy with uniform loss rate $\gamma/2$ while intra-cell couplings are {non-reciprocal}. The bare photonic lattice is a non-Hermitian generalization of the SSH model. Each emitter now couples to the lattice at {\it two} different sites whose respective couplings differ by a $\pi/2$ phase. (c): Schematics of the bare field Hamiltonian $H_f$ (odd $N$) under open BCs (open loop) and the corresponding induced effective Hamiltonian of the emitters, $H_{\rm eff}$ (closed loop) for $N_e=N$, and $\gamma=2 J$. Both Hamiltonians feature fully non-reciprocal couplings but with opposite chirality, where $H_{\rm eff}$ in particular implements a dissipative Hatano-Nelson model. Remarkably, $H_{\rm eff}$ is translationally invariant despite the bare field (hence the total system) breaks translational invariance.  \label{setup}}
		\end{figure}

		The setup we consider [see \fig\ref{setup}(a)] comprises a composite 1D photonic lattice (coupled-cavity array), whose unit cell consists of a pair of cavities denoted with $a$ and $b$. 
		\begin{figure*}
			\centering
			\includegraphics[width=17.5cm]{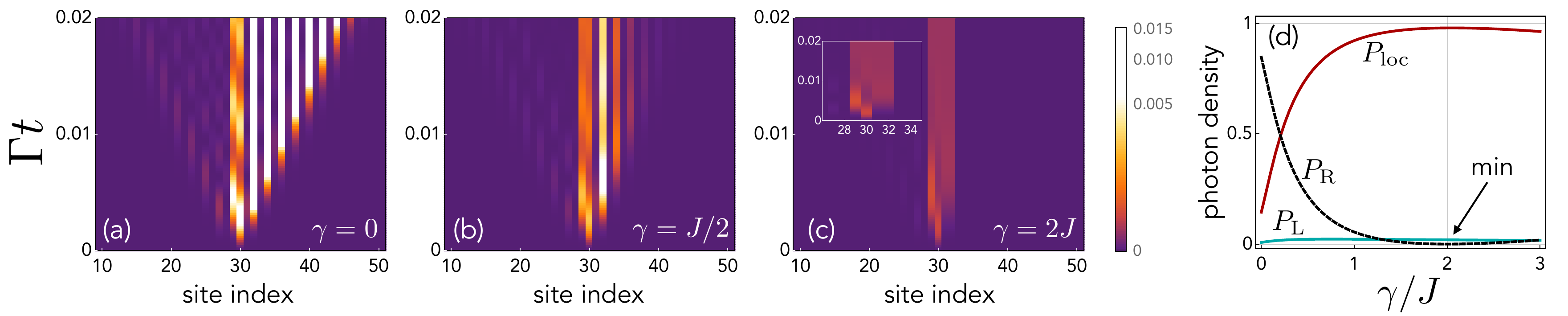}
			\caption{{\it Field dynamics during spontaneous emission}. (a)-(c): Spatial profile of photon density $|\langle \eta_n |\ket{\Psi_t}|^2$ versus time, where $\ket{\Psi_t}=e^{-i H t}\ket{\Psi_0}$, $\ket{\eta_n}=\ket{g}\eta_n^\dag \ket{\rm vac}$, $\eta=a,b$ [referring to the original picture of \fig\ref{setup}(a)]. In the plots, we re-indexed cavities in a way that neutral (lossy) cavities are labeled by odd (even) site indexes. We set $g=0.1 J$ and $N=100$ with the atom coupled to the lossy cavity of cell $n=15$ [see \fig\ref{setup}(a)]. Time is measured in units of $\Gamma^{-1}$ with $\Gamma=g^2/(4J)$. The atom's excited-state population $p_e=|\langle e| \langle{\rm vac}|\ket{\Psi_t}|^2$ decays exponentially as $p_e(t)=e^{-\Gamma t}$. (d): Functional dependence of $P_{\rm loc}$, $P_R$ and $P_L$ on the loss rate $\gamma/J$ (where each probability is rescaled to the sum $P_{L}+P_{\rm loc}+P_{\rm R}$). Here, $P_{\rm loc}$ is the time-averaged probability to find the photon in the cell where the atom lies or the right nearest-neighbour cell (four cavities overall), while $P_L$ ($P_R$) is the probability to find it in the remaining left (right) part of the lattice. We set an average time $t_{\rm av}\sim 20 J^{-1}$ with $g$ small enough such that $t_{\rm av}<\Gamma^{-1}$.} \label{propagation}
		\end{figure*}
		Importantly, only cavities $b$ are leaky the associated loss rate being $\gamma$. 
		By denoting with $a_n$ ($b_n$) the bosonic annihilation operator of cavity $a$ ($b$) in the $n$th cell, the bare Hamiltonian of the field reads (we set $\hbar=1$ throughout)
		\begin{align}
			H_f=\frac{J}{2}\sum_{n=1}^N  &\left[ a_{n}^\dag b_{n+1}+ b_{n}^\dag a_{n+1}-i a_{n}^\dag a_{n+1}+i b_{n}^\dag b_{n+1}\right.\nonumber\\
			&\left.+ 2a_{n}^\dag b_{n}+{\rm H.c.}\right]-i\gamma\sum_{n=1}^N b_{n}^\dag b_{n}\,,\label{Hf}
		\end{align}
		with $N$ the numbers of lattice cells.
		The first line describes the interaction between neighbouring cells, \ie the $a$-$a$ and $b$-$b$ horizontal couplings and the $a$-$b$ diagonal couplings with strength $J/2$ [see \fig\ref{setup}(a)]. In the second line, the first term describes the intra-cell interaction, \ie the vertical $a$-$b$ couplings (strength $J$), whereas the last term accounts for the local losses on $b$ cavities. Note that for $\gamma=0$ we would have $H_f^\dag=H_f$, namely the non-Hermitian nature of the field Hamiltonian comes only from the local losses on $b$ cavities (the overall setup being passive). Model \eqref{Hf} is well-known in the non-Hermitian physics literature as \emph{Lee model}~\cite{LeePRL2016,BergholtzRMP21}.
		
		The system additionally comprises $N_e$ identical two-level quantum emitters (``atoms"), each locally coupled under the rotating wave approximation to a lossy cavity $b$ [see \fig\ref{setup}(a) showing the case $N_e=N$]. The total Hamiltonian is thus 
		\begin{equation}
			H=H_f+\sum_{i=1}^{N_e} g\,(\sigma_{i}^\dag b_{n_i}+b_{n_i}^\dag \sigma_{i})\label{H1}
		\end{equation}
		with $n_i$ the cavity directly coupled to the $i$th atom and where $\sigma_i=\ket{g}_i\!\bra{e}$ is the pseudo-spin ladder operator of the $i$th atom with $\ket{g}$ and $\ket{e}$ respectively the ground and excited states. 
		
		We anticipate that the physical properties which we are going to focus on involve only a single excitation and are thus insensitive to the nature of the ladder operators $\sigma_{i}$ of the emitters, which could thus be thought as cavities/oscillators themselves~\cite{LonghiPRB2009,CrespiPRL19}. Our system  could thus be implemented as well in an all-photonic scenario.
		
		In the above, we assumed that the cavities (either $a$ or $b$) and emitters have all the same frequency $\omega_0$ and set this to zero (\ie energies are measured from $\omega_0$). 
		
		A key feature of the bare photonic lattice [\cf \fig\ref{setup}(a) and Hamiltonian $H_f$] is that, for $\gamma\neq0$, it is {\it non-reciprocal}\, in that photons propagate preferably from right to left. Thus losses endow the structure with an intrinsic left-right asymmetry. 
		One can show that the complex $a-a$ couplings energetically favour left propagating photons and the $b-b$ couplings favour right propagating ones. 
		Indeed, under the standard Peierls substitution (see \eg \rrefs \cite{FeynmanLec}), the kinetic energy associated to a hopping term is minimized by the momentum $k=\theta$, where $\theta$ is the complex phase of the hopping amplitude, which is $\theta=-\pi/2$ for the $a-a$ couplings and $\theta=\pi/2$ for the $b-b$ couplings.
		When losses are present (\ie for $\gamma\neq0$) the left-right symmetry is broken because right-propagating photons (lying predominantly on $b$ sites) are more subject to dissipation than left-propagating ones. This effectively results in photons propagating leftwards with higher probability than rightwards.
		
		Such a dissipation-induced non-reciprocity, which was shown also in other lattices (see \eg \rref{}\cite{LonghiSciRep2015}), can be formally derived by performing the field transformation \cite{BergholtzRMP21} $\{a_n, b_n\}\rightarrow \{ \alpha_n,\beta_n\}$ with
		\begin{equation}
			a_n = \tfrac{1}{\sqrt{2}}(\alpha_n-i \beta_n)\,,\,\,\, b_n = -\tfrac{i}{\sqrt{2}}(\alpha_n+i \beta_n)\,.\label{transformation}
		\end{equation}
		This unitary, which is local in that it mixes cavity modes of the same cell, defines a new picture where the free field Hamiltonian now reads [see \fig\ref{setup}(b)]
		\begin{align}\label{H-SSH}
			H'_f&=
			\sum_{n}\left[
			\left(J+\frac{\gamma}{2}\right)\alpha_n^\dagger \beta_n
			+
			\left(J-\frac{\gamma}{2}\right)\beta_n^\dagger \alpha_n\right]\\
			&+\sum_n 
			J\,(\alpha_{n+1}^\dagger \beta_n + {\rm H.c.})-i\frac{\gamma}{2}\sum_n (\alpha_n^\dagger \alpha_n+\beta_n^\dagger \beta_n).\nonumber\,\,\,\,\,\,\,\,
		\end{align}

		This tight-binding Hamiltonian is a non-Hermitian generalization of the Su-Schrieffer–Heeger (SSH) model \cite{SSH_PRL,BergholtzRMP21}. Unlike the original picture, $H'_f$ features uniform loss on all cavities with rate $\gamma/2$. Remarkably, intra-cell couplings are now manifestly {\it non-reciprocal} for non-zero $\gamma$: the hopping rate of a photon from site $\alpha_n$ to $\beta_n$ differs from that from $\beta_n$ to $\alpha_n$ (respectively $J+\tfrac{\gamma}{2}$ and $J-\tfrac{\gamma}{2}$). Inter-cell couplings $J$ are instead reciprocal. We see that, whenever $\gamma\neq 0$ [non-zero cavity leakage in the original picture, see \fig\ref{setup}(a)] the mapped lattice features an intrinsic {chirality} (\ie non-reciprocity) in that the rate of photon hopping depends on the direction (rightward or leftward). At the critical value $\gamma=2J$, which corresponds to an exceptional point (EP) of the bare lattice \cite{LeePRL2016}, the intra-cell couplings are fully non-reciprocal (all couplings $\alpha_n\rightarrow \beta_n$ vanish). Thus at this EP photons can only propagate to the left.
		
		Consider now the {\it total} Hamiltonian in the new picture, which using \eqref{transformation} reads [\cf\eqs(\ref{H1}) and (\ref{H-SSH})]
		\begin{equation}
			H'=H'_f+\sum_{i=1}^{N_e} \frac{g}{\sqrt{2}}\,\left(\sigma_{i}^\dag (\beta_{n_i}-i  \alpha_{n_i})+{\rm H.c.}\right)\,.\label{H2}
		\end{equation}
		Notably [see \fig\ref{setup}(b)] in the new picture the atom-field interaction is no longer local as each atom is coupled to {\it both} cavities $\alpha$ and $\beta$ of the same cell. The corresponding (complex) couplings have the same strength but, importantly, a $\pi/2$ phase difference.

		Thus, to sum up, in the picture defined by \eqref{transformation}, the system features: (i) uniform losses, (ii) intra-cell non-reciprocal photon hopping rates and (iii) bi-local emitter-lattice coupling. The simultaneous presence of these three factors  is key to the occurrence of the phenomena to be presented.

		\section{Spontaneous emission of one emitter}
		To begin with, we consider only one emitter ($N_e=1$) and study spontaneous emission (initial joint state $\ket{\Psi_0}=\ket{e}\ket{\rm vac}$ with $\ket{\rm vac}$ the field's vacuum state) and we set $g\ll J$. When $\gamma=0$ (no loss), the bare lattice is effectively equivalent to a standard tight-binding model with uniform nearest-neighbour couplings [see \fig\ref{setup}(b)] yielding a single frequency band of width $2J$ with the atom's frequency at its center. 
		
		\figs\ref{propagation}(a)-(c) report the time behavior of the photon density profile across the lattice for different loss rates $\gamma$, while the atom's excited-state population decays exponentially as $p_e=e^{-\Gamma t}$ with $\Gamma=g^2/(4J)$ (not shown in the figure; see caption for details).
		
		For $\gamma=0$ (no loss), directional emission occurs in that the photon propagates predominantly to the right. This is a known effect \cite{RamosPRA16} due to the effective bi-local coupling and $\pi/2$ phase difference in the picture of \fig\ref{setup}(b), which effectively suppresses the interaction of the emitter with left-going modes of the field.
		As $\gamma$ is turned on (lattice leaky) the behaviour considerably changes [see \figs\ref{propagation}(b)-(c)]. Based on the previously discussed non-reciprocity of intra-cell couplings [see \fig\ref{setup}(b)], one might now expect the emitted photon to propagate away mostly to the left (in contrast to the $\gamma=0$ case). Instead, this behaviour is generally exhibited only by a tiny fraction of emitted light.
		Rather, a significant part {\it localises} within a very narrow region of the lattice and eventually 
		leaks out on a long time scale of the order of $\Gamma^{-1}\gg \gamma^{-1}$. Such photon localisation dominates for $\gamma=2J$ [see \fig\ref{setup}(c)], at which value it occurs strictly in two cells only: the one directly coupled to the atom and the nearest neighbour on the right. This is best illustrated in \fig\ref{propagation}(d), where the time-averaged fraction of light localisation in these two cells ($P_{\rm loc}$) is plotted versus $\gamma/J$ along with the fraction lying in the remaining left and right part of the lattice ($P_{L}$ and $P_R$, respectively). We note that $P_{\rm loc}$ is maximum at the EP, where $P_R=0$ and $P_L\simeq 0$ (for $g\rightarrow 0$, $P_L\rightarrow 0$).
		\section{Many emitters}
		We consider next a pair of quantum emitters and study the (dissipative) dynamics of excitation transfer between them when one is initially excited and the other is in the ground state. We again set $\gamma=2J$ [see \fig\ref{setup}(b)], hence the photonic lattice has an intrinsic {\it leftward} chirality.
		\begin{figure}[t]
			\centering
			\includegraphics[width=8.cm]{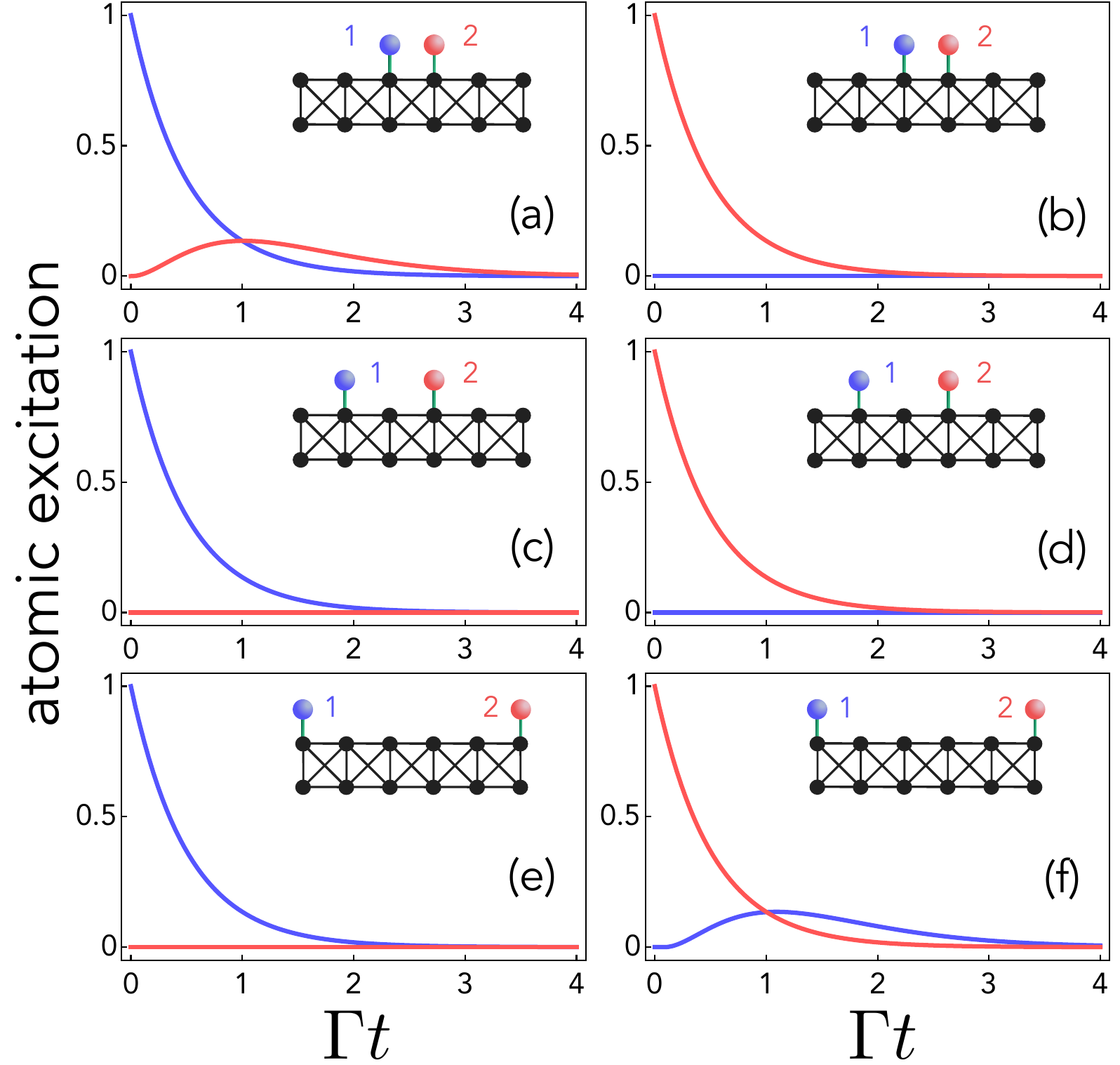}
			\caption{{\it Excitation transfer between two quantum emitters}. We consider two quantum emitters ($N_e=2$) and set $J=\gamma/2$. We plot the time behavior of the emitter 1's excited-state probability $p_1$ (blue line) and that of emitter 2, $p_2$ (red) for the initial state $\ket{\Psi_0}=\ket{e}_1\ket{g}_2\!\ket{\rm vac}$ [panels (a), (c), (e)] and $\ket{\Psi_0}=\ket{g}_1\ket{e}_2\!\ket{\rm vac}$ [panels (b), (d), (f)], where $p_1=|_1\!\langle e| \,_2\!\bra{g}\!\langle{\rm vac}\ket{\Psi_t}|^2$ and an analogous definition holds for $p_2$. The inset in each panel shows the cells where the emitters sit in: nearest-neighbour cells [panels (a) and (b)], non-nearest-neighbour cells in the bulk [panels (c) and (d)], edge cells [panels (e) and (f)]. We set $g=0.1J$. \label{transf}}
		\end{figure}
		When the atoms lie in nearest-neighbour cells [see \fig\ref{transf}(a)], an excitation initially on the left emitter is partially transferred to the right emitter with a characteristic rate $\sim \!\Gamma$ with both emitters eventually decaying to the ground state (transfer is only partial because of the leakage). Notably, as shown by \fig\ref{transf}(b), the reverse process does not occur: if the excitation now sits on the right emitter, this simply decays to the ground state with the left atom remaining unexcited all the time. Thus the field mediates a fully {\it non-reciprocal} (dissipative) interaction between the emitters. At first sight, one might expect this second-order interaction to straightforwardly follow from the aforementioned intrinsic uni-directionality of the bare lattice [recall \fig\ref{propagation}(b) for $\gamma=2J$]. Yet, note that the directionality resulting from \figs\ref{transf}(a) and (b) is {\it rightward} in contrast to that of the lattice which, as said, is leftward [\cf \fig\ref{propagation}(b)]. Later on, we will show that the lattice unidirectionality is indeed a key ingredient for such a non-reciprocal atomic crosstalk, but -- notably -- not the only one. 
		
		Besides being non-reciprocal, the atom-atom effective interaction is {\it exactly} limited to emitters sitting in {\it nearest-neighbour} cells. This can be checked [see \figs\ref{transf}(c) and (d)] by placing the emitters in any pair of {\it non}-nearest-neighbour cells, in which case, no matter what atom is initially excited, no transfer occurs.  A  notable exception to this behaviour yet arises when the lattice is open and emitters sit just on the two opposite edge cells. In this configuration [see \figs\ref{transf}(e) and (f)], counter-intuitively, the coupling is again non-zero and fully non-reciprocal. The associated strength and directionality is just the same (up to a sign) as if the lattice were periodic and the two edge emitters were sitting next to each other [see \figs\ref{transf}(a) and (b)].

\begin{figure}
			\centering
			\includegraphics[width=6.cm]{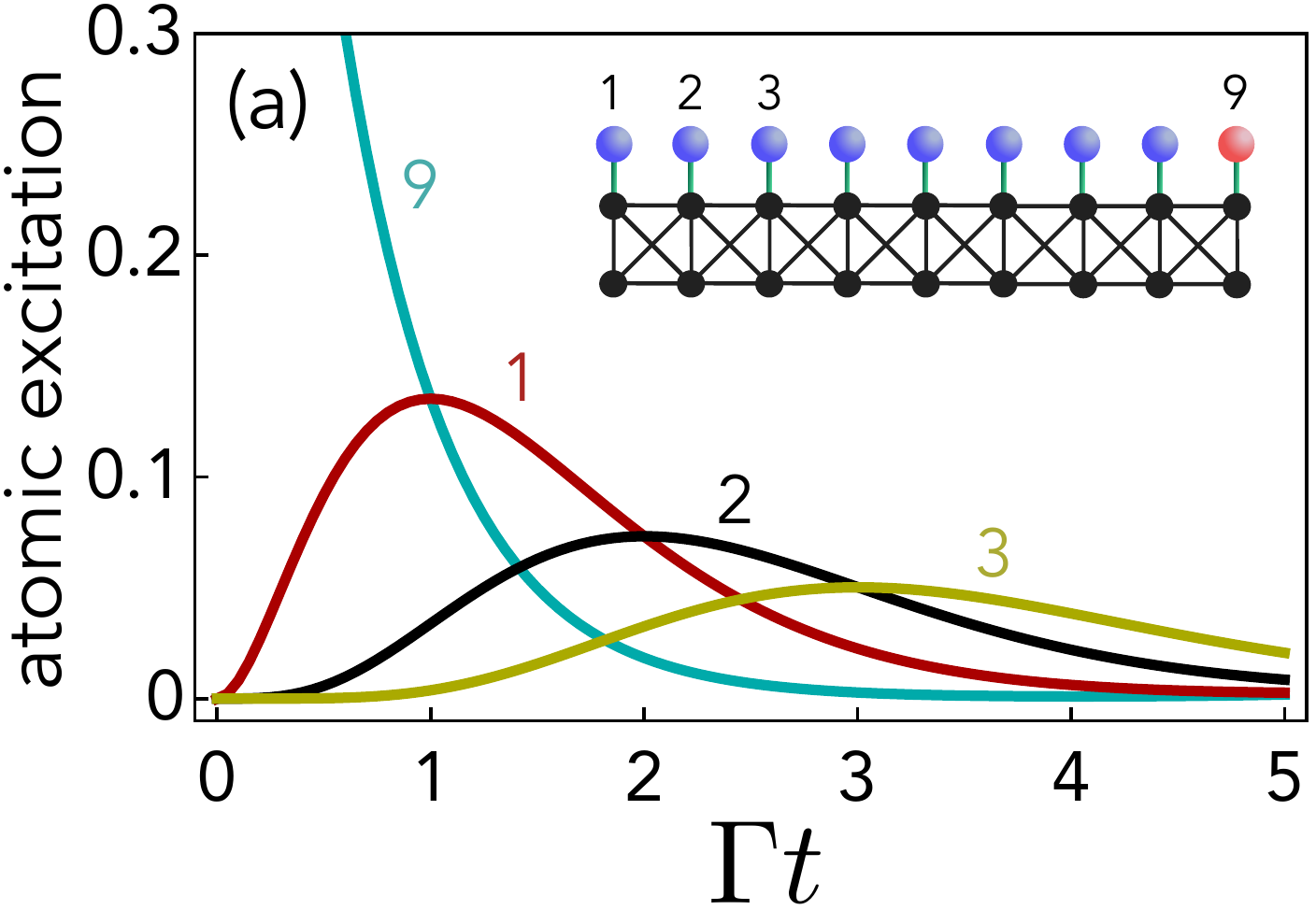}
			\caption{{\it Many-emitter excitation transfer}. We set $N_e=N=9$ (one emitter/cell) and plot against time the excited-state probability of atoms $n=1$ (red line), $n=2$ (black), $n=3$ (yellow) and $n=9$ (cyan) when atom $n=9$ (sitting on the lattice right edge) is initially excited. We set $g=0.1 J$. \label{transfN}}
		\end{figure}

		Results analogous to those in \fig\ref{transf} hold also for many emitters, in particular in the case $N_e=N$ (one atom per unit cell). \fig\ref{transfN} is the $N$-atom analogue of \fig\ref{transf}(f): it clearly shows that an excitation initially on the $N$th atom (on the right edge cell) is first transferred to atom $1$ (sitting on the left edge), then atom 2, then 3 and etc. Again, this behavior is compatible with nearest-neighbour non-reciprocal (rightward) effective couplings between the emitters where -- remarkably -- the emitters on the edges couple to one another as if the lattice were translationally invariant (ring). Indeed, it can be checked that plots in \fig\ref{transfN} remain identical if the lattice is now subject to periodic BCs (no edges).
		\section{Effective Hamiltonian}
\begin{figure}
			\centering
			\includegraphics[width=8.5cm]{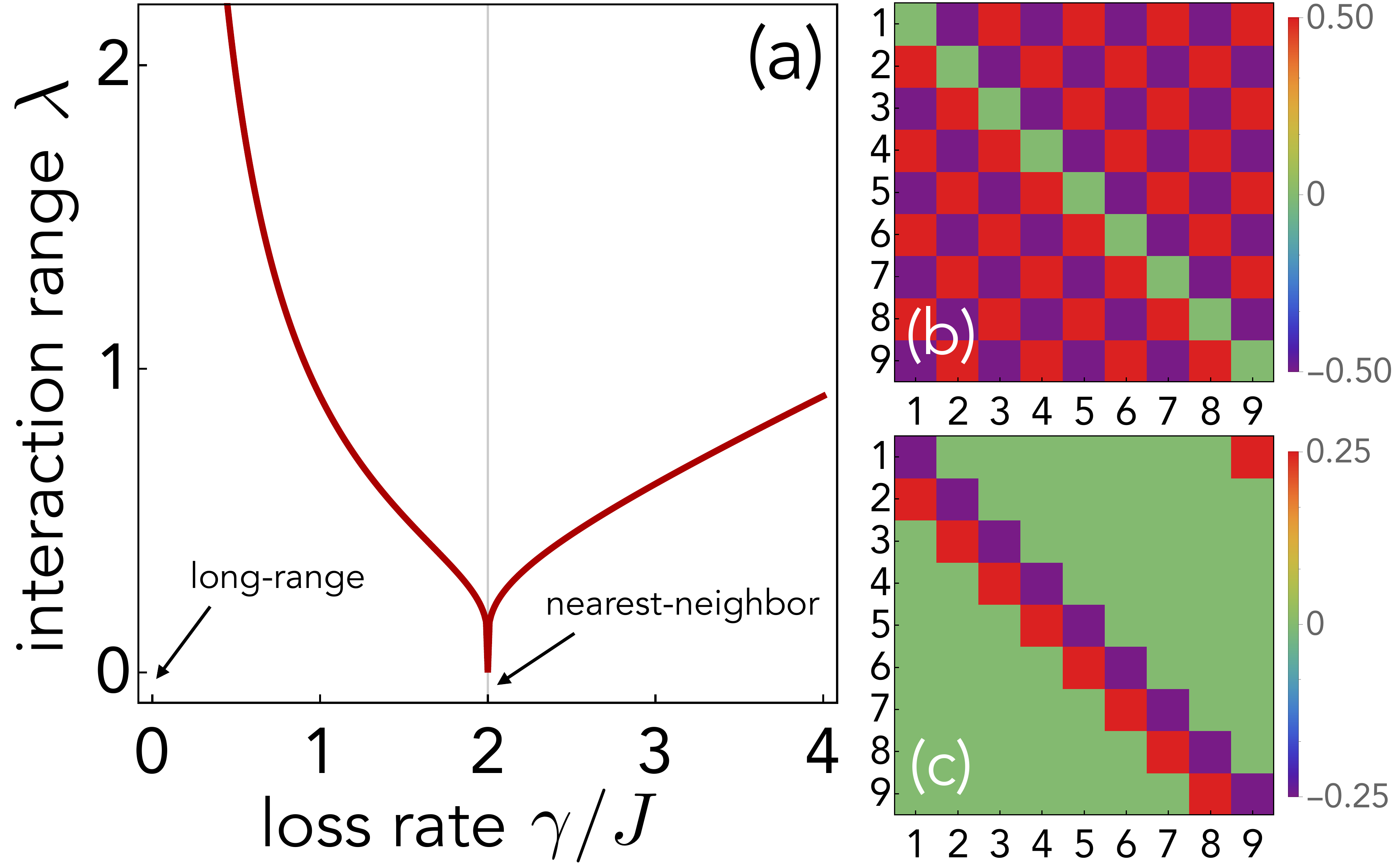}
			\caption{{\it Photon-mediated couplings between emitters}. (a): Interaction range of $H_{\rm eff}$ (see main text) versus $\gamma$. (b) and (c): Matrix plot of ${\cal H}_{mn}$ [\cf\eqref{Hmn}] (imaginary part) for $\gamma=0$ (b) and for $\gamma=2J$ (c) in units of $g^2$ (the real part vanishes). In panel (c), note the upper right corner witnessing that $H_{\rm eff}$ is translationally invariant. All plots are independent of the lattice boundary conditions.\label{figHeff}}
		\end{figure}

		All these dynamics (in particular) are well-described by the effective Hamiltonian of the emitters, which for a bare lattice with periodic BCs reads 
		\begin{align}\label{Heff}
			H_{\rm eff}&\!=\!\sum_{i j} \mathcal{H}_{n_i n_j}\,\sigma_{i}^\dag\sigma_{j}\quad\text{ with }\\
			\mathcal{H}_{m\neq n}&\!=\!i\,4  g^2\! J\,
			\frac{(\gamma\!-\!2J)^{m\!-\!n\!-\!1}}{(\gamma\!+\!2J)^{m\!-\!n\!+\!1}}\,,\,\,  \mathcal{H}_{mm}\!=\!- i \frac{g^2}{\gamma\!+\!2J}\label{Hmn}
		\end{align}
		where periodic BCs are understood, \ie in~\eqref{Hmn} any $n$ is equivalent to $n+N$. Thus $H_{\rm eff}$ is {\it traslationally invariant}.
		This non-Hermitian effective Hamiltonian can be derived analytically in the weak-coupling Markovian regime ($g\ll J$) through a natural non-Hermitian generalization \cite{LonghiPRB2009} of the standard resolvent method \cite{CohenAP,Economou2006,SanchezPRA20}. For $\gamma>0$ and $N\gg \lambda$, where we defined the interaction range $\lambda$ as 
		\begin{equation}
			\lambda^{-1}=-\ln\left|\frac{\gamma-2J}{\gamma+2J}\right|\,,
		\end{equation}
		the entries of ${\cal H}_{mn}$ above the main diagonal vanish (i.e. for $m<n$). Hence, inter-emitter couplings are non-reciprocal with rightward chirality for any $\gamma>0$. Instead, the interaction range $\lambda$ is strongly dependent on $\gamma$ (see \fig\ref{figHeff}). For $\gamma=0$ (no loss) $\lambda$ diverges, witnessing that couplings are purely long-range [see matrix plot in \fig\ref{figHeff}(b)]: all possible pairs of atoms are coupled with the same strength (in modulus) \cite{RamosPRA16} [this can be checked from~\eqref{Hmn} for $\gamma\rightarrow 0$]. As $\gamma$ grows up, the interaction range decreases until vanishing at the lattice EP $\gamma=2 J$ -- where it exhibits a critical behaviour (see cusp) -- and then rises again as $\gamma>2J$. The zero occurs because at $\gamma=2J$ [\cf\eqref{Hmn} for $\gamma\rightarrow 2J$] ${\cal H}_{m>n}$ is non-zero only for $m=n+1$ where it takes the value ${\cal H}_{n+1,n}=i g^2/(4J)=i \Gamma \equiv {\cal H}_{1N}$ [see matrix plot in \fig\ref{figHeff}(c)].
		At this point of the parameter space, therefore, besides being effectively periodic (see above) the non-reciprocal interaction between the emitters is exactly limited to nearest neighbours: this implements a {\it Hatano-Nelson model} \cite{HatanoPRL96} with fully non-reciprocal hopping rates and uniform on-site losses under periodic BCs. The results in \figs\ref{transf} and \ref{transfN} fully reflect these properties.
		
		Even more remarkably and counter-intuitively, it can be demonstrated [see Supplement 1, Section 3] that, for odd $N$, $H_{\rm eff}$ is {\it insensitive to the BCs} of the lattice (matching the results of \fig\ref{transfN}). In other words, even if the lattice is subject to open BCs (ring with a missing cell) $H_{\rm eff}$ is always given by~\eqref{Heff}. For even $N$, the hopping rate across the missing cell is modified by just an extra minus sign [see Supplement 1, Section 3.C]. \fig\ref{setup}(c) sketches the open lattice for $N_e=N$ (one atom/cell) and $\gamma=2J$: both $H_f$ and $H_{\rm eff}$ feature fully-non-reciprocal couplings yet with opposite chirality and, moreover, $H_{\rm eff}$ is periodic while $H_{f}$ is not. We thus in particular get that photons can mediate translationally-invariant interactions between the emitters despite the field (hence the total system) lacking translational invariance.
		
		\section{Atom-photon dressed state}
		Hamiltonian~\eqref{Heff} can be understood in terms of a dressed atom-photon state $\ket{\Psi}$ mediating the 2nd-order interaction between two generic emitters $i$ and $j$ according to the scheme: emitter $i$ $\rightarrow \ket{\Psi}\rightarrow$ emitter $j$, where $\ket{\Psi}$ is a state where $i$ is dressed by a single photon. The resulting $i$-$j$ coupling is non-zero provided that $\ket{\Psi}$ has non-zero amplitude on the location of $j$ [see Supplement 1, Section 3.D]. Similar descriptions were successfully applied to dissipationless interactions for lossless lattices with emitters inside bandgaps \cite{Douglas2015b,Gonzalez-Tudela2015,Gonzalez-Tudela2017a,Gonzalez-Tudela2018a,Bello2019,ZuecoSaw2020,LeonfortePRL2021}, in which case $\ket{\Psi}$ is stationary. In our lossy gapless lattice, instead, interactions between emitters are dissipative and $\ket{\Psi}$ metastable.

		To pinpoint the essential physics, we set $\gamma=2J$ (EP) and consider first an emitter sitting in any {\it bulk} cell indexed by $n
		=\nu$. It is convenient to refer to the picture in \fig\ref{setup}(b) and introduce 
		a light notation such that $\ket{e}\ket{\rm vac}\rightarrow \ket{e}$, while $\ket{g}\ket{\eta_n}\rightarrow\ket{\eta_n}$ with $\eta=\alpha,\beta$.
		\begin{figure}[h!]
			\centering
			\includegraphics[width=8.5cm]{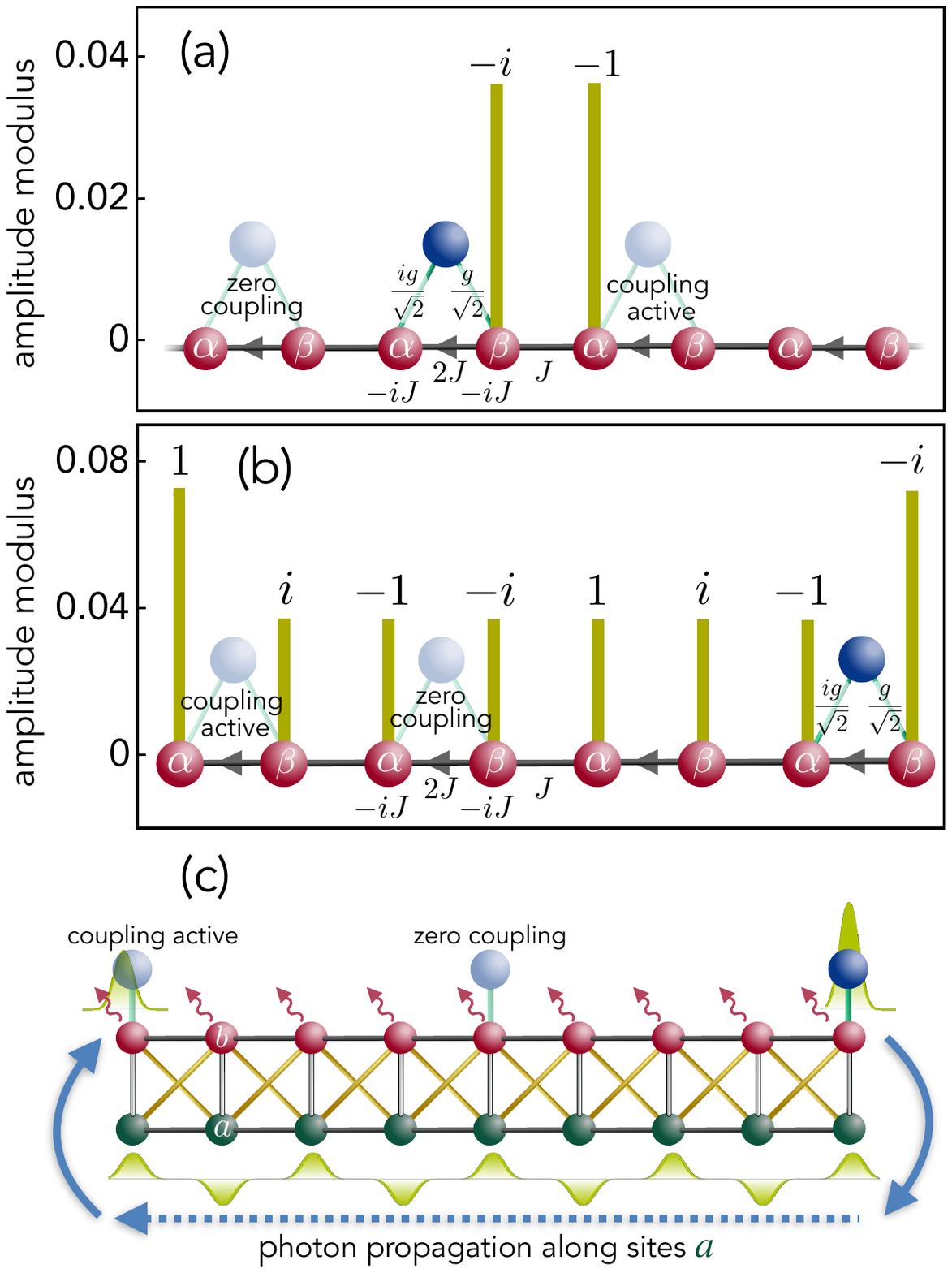}
			\caption{{\it Atom-photon dressed state mediating the emitter-emitter interaction}. (a): Dressed state $\ket{\Psi}$ forming when a quantum emitter (``source") is coupled to a bulk cell. Vertical bars measure the photonic wavefunction modulus. Another emitter (shaded) couples to $\ket{\Psi}$ (``coupling active"), hence to the source emitter, only provided that it sits in the right nearest-neghbour cell. If not (``zero coupling"), the two emitters do not interact. (b): Long-range dressed state arising when the source atom is coupled to the right-edge cell. Due to phase cancellation, an emitter placed in any bulk cell remains uncoupled from $\ket{\Psi}$ unless it lies on the opposite edge. In panels (a) and (b) we set $\gamma=2J$ and referred to the picture in \fig\ref{setup}(b) defined by \eqref{transformation}. (c) Pictorial representation of the state in panel (b) but in the original picture [\cf\fig\ref{setup}(a)]. Note that the photonic wavefunction is zero on all $b$ sites in the bulk, which gives rise to a dissipationless channel bussing excitations between the two edges without involving emitters in the bulk. 
				\label{figBS}
			}
		\end{figure}
		One can check by direct substitution that, to the 2nd order in $g/J$, $H$ admits the eigenstate and associated energy
		\begin{equation}
			\ket{\Psi}=\ket{e}-i\tfrac{g}{\sqrt{2}\gamma}\left(\ket{\beta_\nu}-i\ket{\alpha_{\nu+1}}\right)\,,\,\,\,\epsilon=- i \Gamma\label{BS}
		\end{equation}
		(recall that $\Gamma=g^2/4J$). Note that $\ket{\Psi}$ is normalized to the 2nd order in $g$, while $\ket{\Psi}\rightarrow \ket{e}$ and $\epsilon\rightarrow 0$ for $g\rightarrow 0$. Most remarkably, $\ket{\Psi}$ is strictly localized in only two lattice cells ($\nu$ and $\nu+1$), in particular on cavities $\beta_{\nu}$ and $\alpha_{\nu+1}$. 
		Such strict localization is possible due to a simultaneous ``decoupling" of $\ket{\Psi}$ from the lattice's right branch (sites $\beta_{\nu+1}, \alpha_{\nu+2},...$) and left branch (..., $\beta_{\nu-1}$, $\alpha_\nu$). The right-branch decoupling requires $\langle \beta_{\nu+1}|H|\Psi\rangle=0$ so that $\ket{\Psi}$ has a node on $\beta_{\nu+1}$, which is guaranteed by the non-reciprocal leftward nature of intra-cell hopping rates. To get the left-branch decoupling, instead, we must require $\langle \alpha_\nu|H|\Psi\rangle=0$ (so that $\ket{\Psi}$ can have node on $\alpha_\nu$). It is easily seen [see Supplement 1, Section 2] that this condition can only be met provided that $\epsilon=-i \Gamma$ (showing the metastable nature of the state) plus, crucially, $\langle \alpha_{\nu+1}|\Psi\rangle\neq 0$. The latter circumstance clarifies why the emitter can couple to another emitter sitting in cell $\nu+1$: any other location will give zero coupling since $\ket{\Psi}$ vanishes everywhere outside cells $\nu$ and $\nu+1$. This explains both the non-reciprocal and nearest-neighbour nature of $H_{\rm eff}$ at the lattice EP [\cf\eqref{Heff}] for atoms in the bulk.
		
		We next consider an open lattice with the source atom now sitting in the cell on the right edge [see \fig\ref{figBS}(b)]. 
		Again by direct substitution, $H$ can be shown to admit the eigenstate 
		\begin{align}
			|\Psi\rangle = 
			\ket{e} 
			-\frac{g}{\sqrt{2}\gamma}
			\sum_{n=1}^{N}	e^{i\pi n} & [ \left(1+\delta_{n1}\right)
			\ket{\alpha_n} \nonumber\\
			&-i
			\left(1+\delta_{nN}\right)\ket{\beta_n}]\label{LS}
		\end{align}
		the associated energy still being $\epsilon=-i \Gamma$. 
		This state is normalized to leading order in $g$ under the condition $g\ll \gamma /\sqrt{N}$ [in line with the Markovian regime assumed to derive~\eqref{Heff}].
		Unlike~\eqref{BS}, $\ket{\Psi}$ is extended across the entire lattice [see \fig\ref{figBS}(b)]. In the bulk sites, the photonic wavefunction has flat modulus but non-uniform phase. Remarkably, the pattern of phases combines with the bi-local nature of emitter-field coupling [see \figs\ref{setup}(b) and \ref{figBS}(b)] in such a way that, due to phase cancellation, another atom placed in any bulk cell cannot couple to $\ket{\Psi}$ (hence to the source atom). This conclusion yet does not apply to the leftmost cell, where 
		$|\langle \alpha_1|\Psi\rangle|\neq |\langle \beta_1|\Psi\rangle|$: thus atoms placed on opposite edges are able to crosstalk. For odd $N$, the resulting coupling strength matches that for nearest-neighbour emitters in the case of \fig\ref{figBS}(a) [see Supplement 1, Section 3.C].
		
		To better grasp the physical mechanism enabling \eqref{LS} to mediate an interaction between edge emitters, it is useful rewriting state \eqref{LS} in the original picture [\cf\fig\ref{setup}(a)] through the inverse of \eqref{transf}. 
		This reads
		\begin{align}
			|\Psi\rangle = |e\rangle - \frac{g}{2\gamma} \sum_{n=1}^N  e^{i\pi n } &[ (2 + \delta_{1n} + \delta_{nN}) |a_n\rangle \nonumber\\
			&+ i(\delta_{1n} - \delta_{nN}) |b_n\rangle ]\,.
		\end{align}
	
		Note that the state has zero amplitude on all the lossy sites $b$ in the lattice {\it bulk}. On the one hand, this explains why state $|\Psi\rangle$ cannot mediate any crosstalk between bulk atoms. On the other hand, it makes intuitive how the mediating photon can bus excitations between the system’s edges without decaying in the bulk: as the unit cell contains a lossless site, there exists a dissipationless channel connecting the two lattice edges as sketched in \fig\ref{figBS}(c).
		
		Finally, we point out that the emergence of states \eqref{BS} and \eqref{LS} relies on the simultaneous occurrence of properties (i)-(iii) at the end of {\it Setup and Hamiltonian}, witnessing  in particular the non-Hermitian nature of the above physics.

		\section{Discussion}
		
		These findings introduce a new quantum optics/photonics paradigm, where ``structured" leakage on the field can shape unprecedented emission properties and second-order emitter-emitter dissipative interactions.		
		Besides engineered leakage, a key ingredient for the predicted physics was shown to be the effectively non-local nature of emitter-field coupling (in a suitable picture). Emitters subject to such unconventional {\it non-local} interaction are dubbed ``giant atoms" in the context of an emerging literature \cite{Kockum5years}. They can be implemented via superconducting qubits \cite{gustafsson2014propagating}, cold atoms \cite{gonzalez2019engineering} or all-photonic setups \cite{longhi2020photonic} and seed tunable dipole-dipole Hamiltonians \cite{KockumPRL2018,kannan2020waveguide, carollo2020mechanism,KockumPRLTopo21}. From such a perspective, the presented results stem from an interesting combination of giant atoms physics, non-Hermitian Hamiltonians and, in some respects, chiral quantum optics \cite{zoller1987quantum, LodahlReviewNature17,DarioTB}, holding the promise for further developments \eg using three-local coupling \cite{KockumPRR20} and 2D non-Hermitian lattices \cite{kremer2019demonstration}.

		We point out that the considered setup [\cf\fig\ref{setup}] is fully passive.  In our framework, this naturally follows from the decay nature of the studied phenomena, a type of non-unitary dynamics currently receiving considerable attention also in other scenarios \cite{sheremet2021waveguide}. On the other hand, the passive nature of our system favours an experimental verification of the predicted dynamics, \eg in photonics (where non-Hermitian Hamiltonians are often implemented through their passive counterparts \cite{Ornigotti_2014}). A circuit-QED  implementation appears viable as well: arrays of resonators coherently coupled to superconducting qubits -- including excitation transfer mediated by atom-photon bound states -- were experimentally demonstrated \cite{Sundaresan2019,painter205,scigliuzzo2021probing} and implementations of lattices like the one in \fig\ref{setup}(a) were put forward \cite{WilsonCreutz99}. 
		Patterned losses can be realized by interspersing resonators with low and high quality factors. This is easily achieved in state-of-the-art settings where external losses can be reduced up to four orders of magnitude compared to photon-hopping rates, while large losses can be obtained and controlled by connecting selectively lattice resonators to transmission lines~\cite{Sundaresan2019,painter205,scigliuzzo2021probing}.

		It is natural to ask whether analogous effects occur also in photonic lattices different from the one considered here.  We checked that this is the case for a sawtooth-like photonic lattice very similar to the one in Refs. \cite {ZuecoSaw2020,lorenzo_intermittent_2021} with added losses on one sublattice. A general classification of the photonic Hamiltonians exhibiting these physical properties is a desirable (and non-trivial) task which is left for future work.

		Finally, whether building on the physics presented here one could realize exotic interactions which are dissipationless (possibly adding active elements) is under ongoing investigation.

		\section*{Acknowledgements}
		
		We acknowledge support from MIUR through project PRIN Project 2017SRN-BRK QUSHIP. AC acknowledges support from the Government of the Russian Federation through Agreement No.~074-02-2018-330 (2). GC acknowledges that results incorporated in this standard have received funding from the European Union Horizon 2020 research and innovation programme under the Marie Sklodowska-Curie grant agreement No 882536 for the project QUANLUX.

		\bibliography{WQEDedge}
		\bibliographystyle{apsrev4-1}
			
		\clearpage
					
		
		\begin{bibunit}
			
			\renewcommand{\bibnumfmt}[1]{[S#1]}
			\renewcommand{\citenumfont}[1]{S#1}
			\renewcommand{\theequation}{S\arabic{equation}}
			\renewcommand{\thefigure}{S\arabic{figure}}
			\renewcommand{\thepage}{S\arabic{page}}  
			\renewcommand{\thesection}{S\arabic{section}}
			\renewcommand{\thetable}{S\arabic{table}}
			\setcounter{equation}{0}
			\setcounter{figure}{0}
			\setcounter{page}{1}
			\setcounter{section}{0}

			\onecolumngrid
			
			\begin{center}
				\textbf{\large Supplementary Information for \\``{Exotic interactions mediated by a non-Hermitian photonic bath}''}\\[.2cm]
				F. Roccati,$^{1}$ S.~Lorenzo,$^{1}$, G.~Calaj{\`o},$^{2}$ G.~M. Palma,$^{1,3}$ A.~Carollo,$^{1,4}$ and F. Ciccarello$^{1,3}$\\[.1cm]
				{\itshape \small ${}^1$Universit$\grave{a}$  degli Studi di Palermo, Dipartimento di Fisica e Chimica -- Emilio Segr$\grave{e}$, via Archirafi 36, I-90123 Palermo, Italy\\
				${}^2$	ICFO-Institut de Ciencies Fotoniques, The Barcelona Institute of\\
				Science and Technology, 08860 Castelldefels (Barcelona), Spain\\
				${}^3$NEST, Istituto Nanoscienze-CNR, Piazza S. Silvestro 12, 56127 Pisa, Italy\\
				${}^4$Radiophysics Department, National Research Lobachevsky State University of Nizhni Novgorod,\\
				23 Gagarin Avenue, Nizhni Novgorod 603950, Russia}\\
				(Dated: \today)\\[1cm]
			\end{center}


			Throughout this Supplementary Information, we refer to the generalized bare-field Hamiltonian
			\begin{align}\label{latticeHam}
				H_f=\sum_n  \left[t_1 a_{n}^\dag b_{n}+\frac{t_2}{2}\left( a_{n}^\dag b_{n+1}+ b_{n}^\dag a_{n+1}-i a_{n}^\dag a_{n+1}+i b_{n}^\dag b_{n+1}\right)+	\T{ H.c.}\right]-i\gamma b_{n}^\dag b_{n}\,,
			\end{align}
			which reduces to the Hamiltonian $H_f$ in the main text when $t_1=t_2=J$. The reasons for this generalization are of a merely technical nature.
			
			\section{Energy spectrum of the bare photonic lattice}

\begin{figure}[h!]
				\centering
				\includegraphics[width=8.5cm]{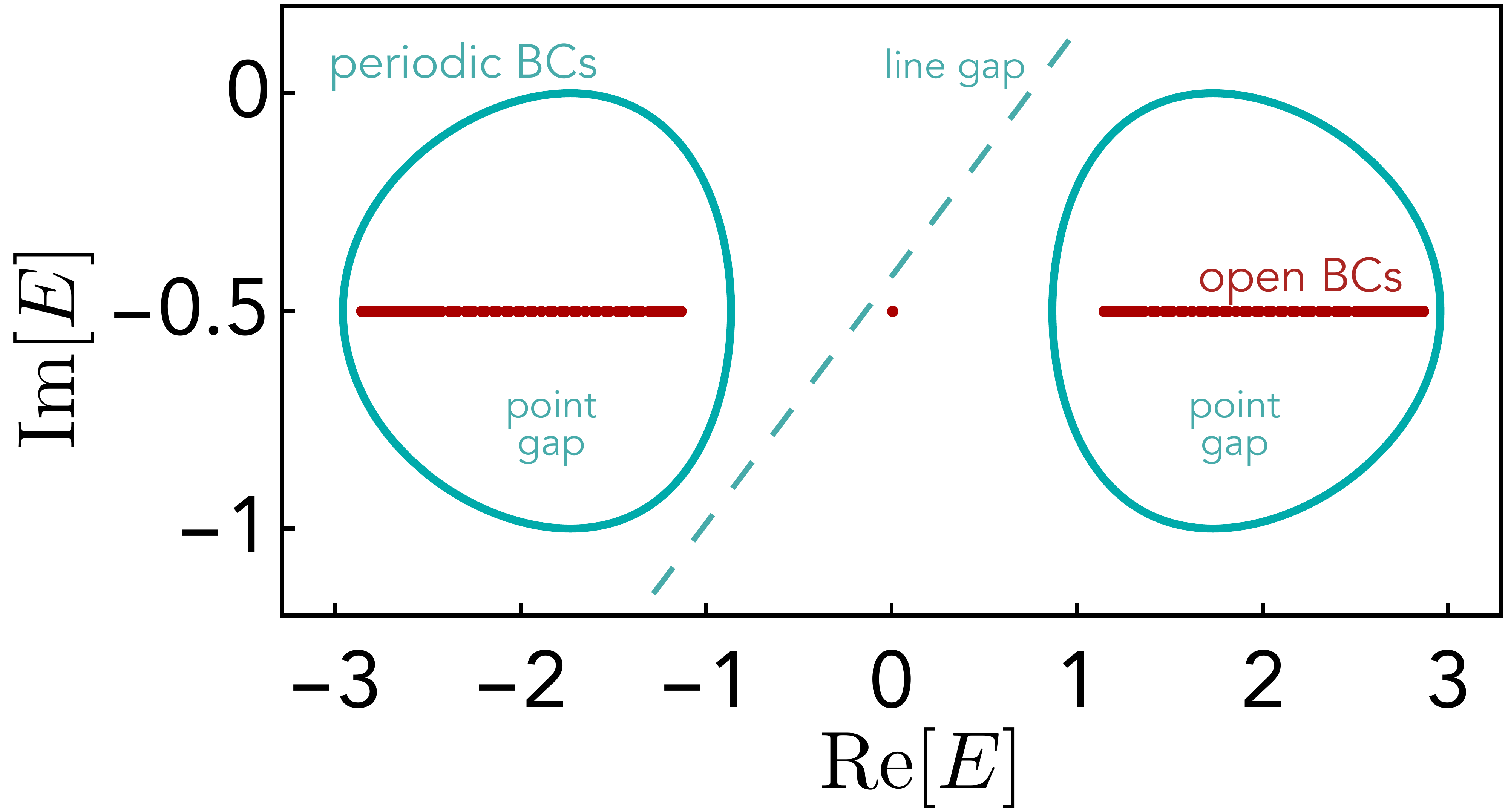}
				\caption{{\it Energy spectrum of the bare lattice.} Under periodic BCs the field energy spectrum can exhibit a line and point gap [$t_2=2t_1$, $\gamma=t_1$] (cyan). High sensitivuty to BCs is manifest in that the spectrum of the field under open BCs (red) is a topologically different curve in the complex $(\T{Re}[E], \T{Im}[E])$ plane.} \label{figSpetrum}
			\end{figure}

			A distinctive and counterintuitive feature of non-Hermitian tight-binding models with non-reciprocal couplings (as our lattice for $\gamma\neq 0$)  is a high sensitivity of the spectrum to boundary conditions (BCs)~\cite{BergholtzRMP21}. Indeed, one can check that under periodic BCs the spectrum of \eqref{latticeHam} is topologically non-trivial exhibiting a line gap (for $t_2\neq t_1$) and a point gap (for $\gamma>0$), see Fig.~\ref{figSpetrum}. On the contrary, the spectrum under open BCs is always trivial in terms of point-gap topology. This discrepancy implies that the appearance of EPs depends on the BCs: under periodic BCs $H_f$ is always diagonalizable, while under open BCs it exhibits an EP at $\gamma=2t_1$ (i.e. at $\gamma=2J$ for $t_1=t_2=J$ as considered in the main text).

			\section{Derivation of the dressed state }

			Consider the emitter in a bulk cell $\nu$ in the picture of \fig1(b) in the main text , the coupling thus being bi-local [condition (iii)]. A dressed state 
			$$\ket{\Psi}=A_e\ket{e}+B_\nu\ket{\beta_\nu}+A_{\nu+1}\ket{\alpha_{\nu+1}}$$ such that $H\ket{\Psi}=\epsilon\ket{\Psi}$ can appear only if: $\gamma=2J$ (we are assuming $t_1=t_2=J$ as in the main text) [condition (ii), i.e.~full non-reciprocity], avoiding amplitudes beyond $\ket{\alpha_{\nu+1}}$ to the right. Imposing $\bra{\alpha_\nu}H\ket{\Psi}=0$, where condition (iii) is crucial, and the Schr\"odinger equation (SE) projected on $\ket{e}$, one gets  $\epsilon=-i\Gamma$ and $A_e=i \, \gamma\sqrt{2}B_\nu/g$. Projecting next the SE on $\ket{\beta_\nu}$ and $\ket{\alpha_{\nu+1}}$, exploiting condition (i) (uniform losses) and in the weak-coupling regime, yields $A_{\nu+1}=-iB_{\nu}$ [see \eq(8) in the main text].
			
			For open BCs of the lattice and an emitter in cell $N$ (right edge), properties (i)-(iii) yield the appearance of an eigenstate at the same energy $\epsilon=-i\Gamma$. By imposing SE, one finds 
			\begin{eqnarray}
				A_n & = &\left(1+\delta_{n,1}\right)e^{i\pi n}A_e/\gamma\sqrt{2}\nonumber\\
				B_n & = &-ig \left(1+\delta_{n,N}\right) e^{i\pi n}A_e/\gamma\sqrt{2}\nonumber
			\end{eqnarray}
			showing that amplitudes on bulk sites are half of those on sites $\alpha_1$ and $\beta_N$ and, most importantly, that there is a $\pi/2$ phase difference between nearest-neighbour sites.

			\section{Effective Hamiltonian}

			\subsection{Derivation  via the resolvent method}
			
			A counter-intuitive property of  $H_{\rm eff}$ is its insensitivity to BCs. We will thus separately consider both periodic and open BCs and show that the resulting effective Hamiltonian, up to a phase factor is indeed the same in the two cases.
			It is quite expected, and indeed well-known, that in many Hermitian short-ranged tight-binding models a change in the BCs does not affect the interaction between emitters when both of these sit far away from the boundaries. However, a striking feature of the present model is that this behaviour occurs even for emitters sitting next to the edges. In the most extreme case, two emitters lying on the two opposite ends of the open lattice interact as if they where sitting next to each other in a lattice subject to periodic BCs.
			Note, however, that this property strictly depends on the weak-coupling approximation, which requires $g\ll J/\sqrt{N}$ (for $t_1=t_2=J$).
			
			In light of the previous discussion,
			in weak-coupling conditions, the effective Hamiltonian $H_{\rm eff}$ of a set of emitters coupled to cavities $b_{n_i}$ and $b_{n_j}$, respectively, can be calculated as~\cite{Bello2019,LonghiPRB2009,CohenAP}
			\begin{align}\label{Solz2}
				H_{\rm eff}=\sum_{ij}\mathcal{H}_{n_i n_j}\sigma_i^\dagger\sigma_j
			\end{align}
			with
			\begin{equation}\label{matelement}
				\mathcal{H}_{nm} \simeq g^2 \bra{0} b_{m} \hat G(E)b_{n}^\dag\ket{0}
			\end{equation}
			where $E$ is the bare frequency of the emitters and 
			\begin{align}
				\hat G(E):=\frac{1}{E-H_f}
			\end{align}
			is the resolvent operator of the bare lattice (we drop subscript $f$ to simplify the notation). For technical reasons, in the derivation of the $H_{\rm eff}$ we need to keep $t_1\neq t_2$, and we will obtain the case in the main in the limit $t_1\to t_2=J$. 
			We will next derive the effective Hamiltonian first when the lattice is subject to periodic boundary conditions (PBCs). Afterward, we will enforce open boundary conditions (OBCs) by introducing an infinite on-site energy (detuning) in a unit cell of the lattice under PBCs. This effectively removes the cell from the periodic lattice, thus effectively turning the geometry into OBCs. In the remainder, we will set $E=0$ in the calculation of the resolvent (namely we set the energy zero to the emitters' frequency).
			
			\subsection{Calculation of the effective Hamiltonian under PBCs of the lattice}
			
\begin{figure}
				\centering
				\includegraphics[width=6.cm]{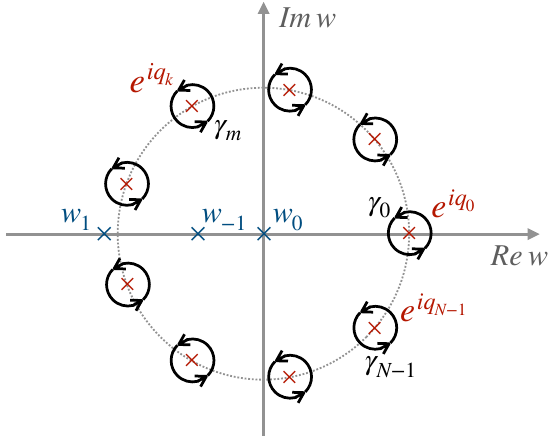}
				\caption{{\it Path of integration-} Sketch of the path of integration for the integral representation~\eqref{PBGreen} of the resolvent $G_{\rm PB}$ under PBCs. The path ${\gamma}=\cup_k\gamma_k$ is the collection of small circles $\gamma_k$ centered on $e^{i q_k}$ with $q_k=\frac{2\pi k}{N}$ and $k=0\dots N-1$. Function $F(w)$ has three poles, $|w_{-1}|< 1$, $w_0=0$ and $|w_1|\ge 1$ which are distinct from all the poles $\{q_k\}_{k=0}^{N-1}$ unless $N$ is an even integer and $t_1=t_2$. For $t_1=t_2$ the pole $w_1=-1$, which for even values of $N$ coincides with $e^{iq_{N/2}}$. In this case, the derivation of $G_{\rm PB}$ needs to be performed with $t_1\neq t_2$ and $H_{\rm eff}$ is eventually calculated in the limit $t_1\to t_2$ (see text).  \label{Fig:poles}}
			\end{figure}

			Due to translational invariance, the lattice Hamiltonian under PBCs can be expressed in its Fourier representation as
			\begin{align}
				H_f^{\rm PB}=\sum_{k=0}^{N-1}\tilde{\Phi}_{q_k}^\dag\cdot \Hk(e^{i q_k})\cdot\tilde{\Phi}_{q_k}\,,\qquad  \Phi_{n}:=\frac{1}{\sqrt{N}}\sum_{j=0}^{N-1}\tilde{\Phi}_{q_k}e^{-iq_k n}\,,\qquad q_k:=\frac{2\pi k}{N}\,. 
			\end{align}
			where $\Phi_{n}:=(a_{n},b_{n})^T$ and 
			\begin{align}
				\Hk (e^{iq})= \begin{pmatrix}
					-t_2\sin{q}&t_1 + t_2\cos{q}  \\
					t_1+t_2\cos{q} &t_2\sin{q}-i\gamma
				\end{pmatrix}
			\end{align}
			is the single particle Hamiltonian in quasi-momentum representation. Correspondingly, the (single-particle) PBCs resolvent in the real-space representation  is given by
			\begin{align}
				G_{\rm PB}(m-n):=-\bra{0}\Phi_{m}\frac{1}{H^{\rm PB}}\Phi_{n}^\dag\ket{0}=-\frac{1}{N}\sum_{j=0}^{N-1}\frac{e^{iq_k(m-n)}}{\Hk(e^{iq_k})}\,.
			\end{align}
			The above expression can be calculated by resorting to the residue theorem as
			\begin{align}\label{PBGreen}
				G_{\rm{PB}}(n)=\frac{1}{2\pi i}\oint_{\gamma}dw\frac{w^n}{w^N-1}F(w)\,,\qquad
				F(w):=-\frac{1}{w\Hk(w)},\qquad\text{ with } e^{iq}\to w\in \mathbb{C}
			\end{align}
			where $0\le n<N$ and the path of integration ${\gamma}=\cup_k\gamma_k$ is the collection of small circles $\gamma_k$ centered on $e^{i q_k}$~(\footnote{If $f(w)$ is analytic in $e^{iq_k}$, one can easily show that 
				\begin{align}
					\frac{1}{2\pi i}\oint_{\gamma_m} \frac{dw}{w} \frac{f(w)}{w^N-1}=\res{e^{iq_k}}\frac{1}{w} \frac{f(w)}{w^N-1}=\lim_{w\to e^{iq_k}}\frac{1}{w} \frac{{w-e^{iq_k}}}{w^N-1}f(w)=\lim_{u\to 1}\frac{1}{u} \frac{{u-1}}{u^N-1}f(u e^{iq_k})=\lim_{u\to 1}\sum_{q=0}^{N-1}u^q f(u e^{iq_k})=Lf(e^{iq_k})	
			\end{align}}) and it is crucial that none of the poles of $F(w)$ coincides with $q_k$ for $k=0\dots N-1$ (see Fig.~\ref{Fig:poles}).
			The expression for $F(w)$ is 
			\begin{align}\label{eq:GwSE2}
				F(w)=\frac{W(w)}{P(w)}\qquad W(w):=\begin{pmatrix}
					i\gamma w-i\frac{t_2}{2}(w^2-1)&t_1 w +\frac{t_2}{2}(w^2+1)  \\
					t_1 w + \frac{t_2}{2}(w^2+1)& i\frac{t_2}{2}(w^2-1)
				\end{pmatrix},
			\end{align}
			where $P[w]:=\Det[-w\Hk(w)]=a w(w-w_{1})(w-w_{-1})$ and $a=-t_2\left(t_1+\frac{\gamma}{2}\right)$, where
			\begin{align}\label{poles}
				w_0=0\,,\qquad	w_{\pm 1}=-\frac{t_1^2+t_2^2 \pm\sqrt{\Delta}}{t_2\left(2t_1+\gamma\right)} \qquad\text{ with }\qquad \Delta=(t_1^2-t_2^2)^2+\gamma^2 t_2^2\,,
			\end{align}
			are the roots of $P(w)$. The roots $\{w_j\}_{j=-1}^1$ are the poles of $F(w)$ whose residues $G_j=\res{w_j}[F(w)]$ are given by
			\begin{align}\label{residues}
				G_0=\frac{1}{2 t_1-\gamma}\begin{pmatrix}
					i&1 \\
					1&-i
				\end{pmatrix},\qquad G_{\pm 1}=\mp\frac{W(w_{\pm 1})}{w_{\pm 1}\sqrt{\Delta}}\,.
			\end{align}
			For $\gamma>0$ and $t_1>0$, one can check that $|w_{-1}|<1$ and $|w_{1}|\ge 1$, where the latter inequality is saturated only when $t_1=t_2$, in which case
			\begin{align}
				w_{1}\simeq-1\,,\qquad w_{-1}\simeq\Lc\,,\qquad\text{where } \Lc:=\frac{\gamma -2 t}{\gamma+2t} \,.
			\end{align}
			One can exploit the residue theorem again to calculate the integral in \eqref{PBGreen} as the residue at infinity minus the sum of the residues on poles different from $e^{i q_k}$, i.e.
			\begin{align}\label{IntPoles3}
				G_{\rm{PB}}(n):=-\res{\infty}\left[\frac{w^{n}}{w^N-1}F(w)\right]-\sum_{j=-1}^{1}\res{w_j}\left[\frac{w^{n}}{w^N-1}F(w)\right]\,.
			\end{align}
			For $n<N$, the residue at infinity vanishes, since
			\begin{align}
				\res{\infty} \left[F(w) \frac{w^n}{w^N-1}\right]=-\res{0} \left[ w^{-2}F(w^{-1}) \frac{w^{-n}}{w^{-N}-1}\right]
				=\res{\infty}[F(w)] \lim_{w\to 0}\frac{w^{N-n}}{1-w^N} =0\,,
			\end{align}
			where we used the fact that $\res{\infty}{F(w)}<\infty$, since $F(w)$ is analytic except for a finite number of isolated poles.
			\emph{The PBCs resolvent in real-space representation} is then given by
			\begin{align}\label{GPB}
				G_{\rm{PB}}(n)=\sum_{j} \frac{w_j^n}{1-w_j^N}G_j \qquad 0\le n<N\,,
			\end{align}
			where $G_j$ are the residues of $F(w)$ given in \eqref{residues}.

			For the case $t_1\simeq t_2$, we define $J$ and $\delta$ such that $t_1=J(1-\delta)$ and $t_2=J(1+\delta)$. Notice that as $\delta\to 0$, the factor $\frac{1}{1-w_{1}^{N}}$ behaves differently with even and odd number of unit cells, with $\frac{1}{1-w_1^{N}}\sim-\frac{\gamma}{4 N J \delta^2}$ for $N$ even and $\frac{1}{1-w_1^{N}}\sim\frac{1}{2}-\frac{N t \delta^2}{\gamma}$ for $N$ odd. 
			Indeed, we will calculate $G_{\rm PB}(n)$ for these two cases separately.
			\paragraph{$N$ odd.--}As noticed, \emph{for $N$ odd} $\frac{1}{1-w_{1}^{N}}\sim\frac{1}{2}-\frac{N t \delta^2}{\gamma}$. By inserting~\eqref{residues},~\eqref{poles} and~\eqref{eq:GwSE2} into \eqref{GPB} we obtain the explicit expression of the resolvent $G_{\rm PB}$, which in the limit $\delta\to 0$ reads
			\begin{align}\label{GPBodd}
				\lim_{\delta\to 0}G_{\rm{\rm PB}}(n)=\begin{cases}
					\frac{(-1)^{\bar{n}}}{2J}\begin{pmatrix}
						i&0  \\
						0&0
					\end{pmatrix}
					+\frac{\Lc^{\bar{n}-1}}{1-\Lc^N}
					\frac{2}{(\gamma+2J)^2}
					\begin{pmatrix}
						-i\frac{\gamma^2}{2J}&-\gamma  \\
						-\gamma&i2J
					\end{pmatrix} 
					& \bar{n}\neq 0\,,\\
					\frac{1}{2J}\begin{pmatrix}
						i&0  \\
						0&0
					\end{pmatrix}+\frac{1}{1-\Lc^N}
					\frac{2}{\gamma^2-4t^2}
					\begin{pmatrix}
						-i\frac{\gamma^2}{2J}&-\gamma  \\
						-\gamma&i2J
					\end{pmatrix} 
					+\frac{1}{\gamma-2J}\begin{pmatrix}
						i&1 \\
						1&-i
					\end{pmatrix}& \bar{n}=0\,,
				\end{cases}	\qquad\text{with }\qquad \bar{n}:=n\!\!\pmod{N}\,.
			\end{align}
			For $\gamma>0$, with moderately large values of $N$ ($\delta^{-1}\gg N\gg -1/\ln|\Lc|$ ), the $G_{\rm{PB}}(n)$ can be formulated in a way which is manifestly chiral, i.e.
			\begin{align}\label{GPBoddTh}
				\lim_{N\to\infty}\lim_{\delta\to 0}G_{\rm{PB}}(n)=
				\begin{cases}
					\frac{(-1)^{n}}{2J}\begin{pmatrix}
						i&0  \\
						0&0
					\end{pmatrix}
					+
					\frac{2\Lc^{\bar{n}-1}}{(\gamma+2J)^2}
					\begin{pmatrix}
						-i\frac{\gamma^2}{2J}&-\gamma  \\
						-\gamma&i2J
					\end{pmatrix} 
					& n> 0\,,\\
					-\frac{1}{\gamma+2J}\begin{pmatrix}
						i\frac{\gamma}{2J}&1 \\
						1&i
					\end{pmatrix}& n=0\\
					-\frac{(-1)^{n}}{2J}\begin{pmatrix}
						i&0  \\
						0&0
					\end{pmatrix}
					& n< 0\,
				\end{cases}	
			\end{align}
			where we relabelled  $n=-\frac{N-1}{2}\dots \frac{N-1}{2}$ and neglected contributions smaller than $\Lc^{N/2}$. 
			\\
			\\
			\paragraph{$N$ even.---}For $N$ even, we notice that $\frac{1}{1-w_1^{N}}\sim-\frac{\gamma}{4 N J \delta^2}$. Indeed, the divergence as $\delta\to 0$ is related to a technical assumption in the derivation of $G_{\rm PB}$ which prescribes that none of the poles $w_{\pm 1}$ should coincide with $e^{iq_k}$ for $m=0\dots N-1$. This assumption fails for even values of $N$ and $t_1=t_2$. Indeed, $w_1=-1$ for $t_1=t_2$ and $e^{iq_{N/2}}=-1$ for even values of $N$. However, we can derive the resolvent $G_{\rm PB}(n)$ for $\delta\neq 0$ and show that the effective Hamiltonian has a finite and meaningful limit for $\delta \to 0$.
			Hence, up to the zero-th order in $\delta$ 
			\begin{align}\label{GPBeven}
				\lim_{\delta\to 0}G_{\rm{\rm PB}}(n)=\begin{cases}
					\frac{(-1)^{\bar{n}}}{N}\begin{pmatrix}
						-i\frac{\gamma}{(2J\delta)^2}+i\frac{N-2\bar{n}}{2J}&\frac{1}{J\delta}  \\
						\frac{1}{J\delta}&\frac{i}{\gamma}
					\end{pmatrix}
					+\frac{\Lc^{\bar{n}-1}}{1-\Lc^N}
					\frac{2}{(\gamma+2J)^2}
					\begin{pmatrix}
						-i\frac{\gamma^2}{2J}&-\gamma  \\
						-\gamma&2it
					\end{pmatrix} 
					& \bar{n}\neq 0\,,\\
					\frac{i}{2N}\begin{pmatrix}
						-i\frac{\gamma}{(2J\delta)^2}+i\frac{N}{2J}&\frac{1}{J\delta}  \\
						\frac{1}{J\delta}&\frac{i}{\gamma}
					\end{pmatrix}+\frac{1}{1-\Lc^N}
					\frac{2}{\gamma^2-4t^2}
					\begin{pmatrix}
						-i\frac{\gamma^2}{2J}&-\gamma  \\
						-\gamma&i2J
					\end{pmatrix} 
					+\frac{1}{\gamma-2J}\begin{pmatrix}
						i&1 \\
						1&-i
					\end{pmatrix}& \bar{n}=0\,,
				\end{cases}
			\end{align}
			Again, for $\gamma>0$ and for sufficiently large values of $N$ ($\delta^{-2}\gg N\gg -1/\ln|\Lc|$ ) we can neglect terms of order $1/N$ and $\Lc^N/2$
			\begin{align}\label{GPBevenTh}
				\lim_{N\to\infty}\lim_{\delta\to 0}G_{\rm{PB}}(n)=
				\begin{cases}
					\frac{(-1)^{n}}{2J}\begin{pmatrix}
						i-i\frac{\gamma}{4JN\delta^2}&0  \\
						0&0
					\end{pmatrix}
					+\frac{2\Lc^{n-1}}{(\gamma+2J)^2}
					\begin{pmatrix}
						-i\frac{\gamma^2}{2J}&-\gamma  \\
						-\gamma&2it
					\end{pmatrix}  
					& n> 0\,,\\
					-\frac{1}{2J+\gamma}\begin{pmatrix}
						i\frac{\gamma}{2J}&1 \\
						1&i
					\end{pmatrix}-\frac{\gamma}{8NJ^2\delta^2}\begin{pmatrix}
						i&0  \\
						0&0
					\end{pmatrix}& n=0\,,\\
					\frac{(-1)^{n}}{2J}\begin{pmatrix}
						-i\frac{\gamma}{4JN\delta^2}&0  \\
						0&0
					\end{pmatrix}
					& n< 0\,.
				\end{cases}	
			\end{align}
			with a similar notation as in \eqref{GPBoddTh}.
			\\
			\\
			\paragraph{Effective Hamiltonian.--}For $\gamma>0$ and for sufficiently large values of $N$ ($\delta^{-2}\gg N\gg -1/\ln|\Lc|$ ) we can neglect terms of order $1/N$ and $\Lc^N/2$ in~\eqref{GPBodd} and~\eqref{GPBeven}, and the effective Hamiltonian can be expressed \emph{for both even and odd values of $N$} as 
			\begin{align}
				\mathcal{H}_{mn}^{\rm PB}=g^2G_{\rm PB}(m-n)_{bb}=ig^2\begin{cases}
					4J\frac{(\Gamma -2 t)^{\overline{m-n}-1}}{(\gamma+2J)^{\overline{m-n}+1}}
					& \overline{m-n}\neq 0\,,\\
					-\frac{1}{\gamma+2J}& \overline{m-n}=0\,,
				\end{cases}	
			\end{align}
			where $\overline{n-m}:=n-m\pmod{N}$. This expression is the same as (7) in the main text. The above expression, can also be cast in a form which is manifestly chiral
			\begin{align}
				\mathcal{H}_{mn}^{\rm PB}=g^2 G_{\rm PB}(m-n)_{bb}=ig^2\begin{cases}
					4J\frac{(\gamma -2 J)^{m-n-1}}{(\gamma+2J)^{m-n+1}}
					& m>n\,,\\
					-\frac{1}{\gamma+2J}& m=n\,,\\
					0& m<n\,,
				\end{cases}	
			\end{align}
			where $-N/2<m-n<N/2$.						
			
			\subsection{Calculation of the effective Hamiltonian under OBCs of the lattice}
			
			As mentioned, an $N$-site lattice with OBCs can be realised by removing a cell from a periodic lattice with $N+1$ sites. Thus, the OBCs Hamiltonian $H_f^{\rm OB}$ can be effectively obtained by adding an infinite on-site energy to a single cell, say $n=0$, to the PBCs Hamiltonian of a $N+1$-site lattice, i.e. 
			\begin{align}
				H_f^{\rm OB}=H_f^{\rm PB}+H_1\quad\text{ with }\quad H_1=\xi\hat{P}\quad\text{ where }\quad\hat{P}=\sum_\ell\Phi_0^\dag\Phi_0\quad\text{ and }\quad \xi\to\infty\,.	
			\end{align}
			Correspondingly, the resolvent operator $\hat G_{\rm OB}(E):=(E-H_f^{\rm OB})^{-1}$ can be obtained non-perturbatively as \cite{Economou2006} 
			\begin{align}\label{OBCG}
				\hat G_{\rm OB}(E)
				=
				\hat G_{\rm PB}(E)-\hat G_{\rm PB}(E)\hat{P}\frac{1}{\hat G_{\rm PB}(E)}\hat{P}\hat G_{\rm PB}(E)\,.
			\end{align}
			We are interested in the case $E=0$, in which case the real-space representation of the resolvent $G_{\rm OB}(m,n):=-\bra{0}\Phi_{m}\frac{1}{H^{\rm OB}}\Phi_{n}^\dag\ket{0}$ takes the form
			\begin{align}\label{OBCSpG}
				G_{\rm OB}(m,n)=G_{\rm{PB}}(m-n)-G_{\rm{PB}}(m)G_{\rm PB}(0)^{-1}G_{\rm{PB}}(-n)\,.
			\end{align}
			Similarly to PBCs, we will calculate the OBCs effective Hamiltonian as $g^2$ times the resolvent operator on the $b$ sublattice [\cf\eqref{Solz2}], i.e. $\mathcal{H}_{\rm nm}^{\rm OB}=g^2G_{\rm OB}(m,n)_{bb}$. For the same technical reasons discussed in the PBCs case, we need to treat even and odd values of $N$ separately.
			\\
			\paragraph{N even.--}Note that an even number of sites $N$ in the OB lattice corresponds to an odd number of sites $N+1$ in the corresponding PB lattice. To evaluate the expression~\eqref{OBCSpG} it is sufficient to consider the contributions of the resolvent $G_{\rm{PB}}(n)$ in the neighbourood of $n=0$, where the potential barrier $H_1$ acts. For simplicity, we will assume $N$ sufficiently large and use the more convenient formulation~\eqref{GPBoddTh} of $G_{\rm{PB}}$ with $N+1$ sites.
			A straightforward calculation leads to
			\begin{align}\label{GOB}
				G_{\rm OB}(m,n)_{bb}&=G_{\rm{PB}}(m-n)_{bb}-\left[G_{\rm{PB}}(m)G_{\rm PB}(0)^{-1}G_{\rm{PB}}(-n)\right]_{bb}
				=\begin{cases} -G_{\rm{PB}}(m-n-1)_{bb}& \text{ for } m>0\wedge n<0\\
					G_{\rm{PB}}(m-n)_{bb}& \text{ otherwise }\,.				
				\end{cases}	
			\end{align}
			Due to the vanishing terms of expression~\eqref{GPBoddTh} for $n<0$, the only non-trivial value of the perturbation $\left[G_{\rm{PB}}(m)G_{\rm PB}(0)^{-1}G_{\rm{PB}}(-n)\right]_{bb}$ may come from the case with $m>0$ and $n<0$. This case corresponds to cells $m$ and $n$ lying on opposite sides of the potential barrier, with $m>n$. Remarkably, the value assumed by $G_{\rm{OB}}(m-n)_{bb}$ exactly coincides up to a phase factor with  $G_{\rm{PB}}(m-n-1)_{bb}$. Notice the correspondence $m-n\leftrightarrow m-n-1$ between the OBCs and PBCs lattice. This correctly accounts for the missing elementary cell which has been effectively removed by the potential barrier $H_1$.  
			\\
			\\
			\paragraph{N odd.}
			Repeating the same calculations for odd values of $N$ and exploiting \eqref{GPBevenTh} with $N+1$ lattice sites yields a very similar result, i.e.
			\begin{align}\label{GOBeven}
				G_{\rm OB}(m,n)_{bb}=
				\begin{cases} 
					G_{\rm{PB}}(m-n-1)_{bb}& \text{ for } m>0\wedge n<0\\
					G_{\rm{PB}}(m-n)_{bb}& \text{ otherwise }\,.				
				\end{cases}	
			\end{align}
			Finally, relabelling the OBCs chain $n \to n-1$ for $n>0$ and collecting the results for odd and even values of $N$ leads to
			\begin{align}\label{HeffOB}
				\mathcal{H}^{\rm{OB}}_{mn}=
				\begin{cases}
					(-1)^{N+1} \mathcal{H}^{\rm{PB}}_{mn}& m>0 \,\wedge\,n<0\,,\\
					\mathcal{H}^{\rm{PB}}_{mn}& \text{otherwise\,.}
				\end{cases}
			\end{align}
			The above expression explicitly demostrates that, for $E=0$, $\gamma>0$, and $t_1=t_2>0$, the effective Hamiltonian with open boundary conditions coincides up to a minus sign with the effective Hamiltonian with periodic boundary conditions. 
			
			As anticipated, a common feature of many Hermitian models with short-range interaction is the insensitivity of the bulk to boundary effects. I.e. for regions sufficiently far away from the borders, most of the dynamical features, such as propagations, interactions between subsystems, etc. are expected to be insensitive to the BCs.
			
			This is not the case here. Remarkably, the effective Hamiltonian of this model displays a striking insensitivity to BCs also in the neighbourhood of the lattice edges. Indeed, \eqref{HeffOB} demonstrates that, even in the extreme case of two emitters coupled to the two opposite ends of the open lattice, these interact to one another as if they where coupled to neighbouring cells in the periodic lattice (up to a sign). 
			\\
			\\
			\paragraph{OB with finite lattice.--}All the arguments above relied on the simplifying assumption that $N$ is sufficiently large. However, even relaxing this condition leads to an expression similar to \eqref{HeffOB} above. For example for $N$ even, using \eqref{GPBodd} with $N+1$ yields
			\begin{align}
				G_{\rm OB}(m,n)_{bb}&=G_{\rm{\rm PB}}(m-n)_{bb}-\left[G_{\rm{\rm PB}}(m)G_{\rm PB}(0)^{-1}G_{\rm{\rm PB}}(-n)\right]_{bb}=-\frac{1-\Lc^{N}}{1+\Lc^{N-1}}G_{\rm{\rm PB}}(m-n-1)_{bb}\,,
			\end{align}
			which shows that the finite-size formula converges exponentially to~\eqref{HeffOB}, i.e.
			\begin{align}
				H_{\rm{eff}}^{\rm{\rm OB}}(m,n)=
				\begin{cases}
					-\frac{1-\Lc^{N}}{1+\Lc^{N-1}}H_{\rm{eff}}^{\rm{\rm PB}}(m-n-1)& m>0\, \wedge \,n<0\,,\\
					H_{\rm{eff}}^{\rm{\rm PB}}(m-n)& \text{otherwise\,.}
				\end{cases}
			\end{align}

			\subsection{Relationship with dressed states}
			
			The biorthogonal completeness relation corresponding to $H_f$ and $H_f^\dag$ (recall that in general $H_f\neq H_f^\dag$) reads
			\begin{equation}
				\sum_k | \psi_k^R \rangle \langle \psi_k^L |= \mathbb{1}
			\end{equation}
			where $\mathbb{1}$ is the identity operator and $\ket{\psi_k^R}$ $\left(\bra{\psi_k^L} \right)$ are right (left) eigenstates of $H_f$. Using this, 
			it is easily shown 
			that 
			the matrix element of the effective Hamiltonian [\cf\eqref{matelement}] can be equivalently rearranged as
			\begin{equation}
				\mathcal H_{\mu\nu} 
				= 
				\bra{0} \sigma_\mu V
				\G(0) 
				V \sigma_\nu^\dagger \ket{0}=g^2
				\bra{b_\mu}
				\G(0) 
				\ket{b_\nu} =
				g
				\braket{b_\mu}{\Psi}\,\label{Hmunu}
			\end{equation}
			(we changed cell indexes as $n\rightarrow\nu$ and $m\rightarrow\mu$ to comply with the notation in Section 6 of the main text).
			Here, $V=H-H_f$ [\cf\eq(2) in the main text] is the atom-field interaction Hamiltonian while $\ket{\Psi}=\ket{e} +g\,\G(0) 
			\ket{b_\nu}$ is the atom-photon dressed state seeded by an atom coupled to the $\nu$th cell. The last step in \eqref{Hmunu} relies on the weak-coupling assumption as in this regime
			$\ket{\Psi} $ [see \eq(8) in the main text], is normalized to the 2nd order in $g$.

			\putbib
		\end{bibunit}

\end{document}